\documentclass[pra,twocolumn,showpacs,floatfix]{revtex4}
\usepackage{graphicx}
\begin{document}

\title{Hysteresis and metastability of Bose-Einstein condensed clouds of atoms
confined in ring potentials}

\author{A. Roussou$^1$, G. D. Tsibidis$^2$, J. Smyrnakis$^3$, M. Magiropoulos$^3$, 
Nikolaos K. Efremidis$^1$, A. D. Jackson$^4$, and G. M. Kavoulakis$^3$}
\affiliation{$^1$Department of Applied Mathematics, University of Crete, GR-71004,
Heraklion, Greece \\
$^2$Institute of Electronic Structure and Laser (IESL), Foundation for Research and 
Technology (FORTH), N. Plastira 100, Vassilika Vouton, 70013, Heraklion, Crete, Greece \\
$^3$Technological Education Institute of Crete, P.O. Box 1939, GR-71004, Heraklion, 
Greece \\
$^4$The Niels Bohr Institute, and Niels Bohr International Academy, Blegdamsvej 17, 
Copenhagen \O, Denmark}

\date{\today}

\begin{abstract}

We consider a Bose-Einstein condensed cloud of atoms which rotate in a toroidal/annular 
potential. Assuming one-dimensional motion, we evaluate the critical frequencies associated 
with the effect of hysteresis and the critical coupling for stability of the persistent 
currents. We perform these calculations using both the mean-field approximation and 
the method of numerical diagonalization of the many-body Hamiltonian which includes 
corrections due to the finiteness of the atom number. 

\end{abstract}
\pacs{05.30.Jp, 03.75.Lm} \maketitle

\section{Introduction}

Numerous experiments on cold atomic gases have been performed in recent years in 
topologically nontrivial trapping potentials, namely in annular/toroidal traps
\cite{Kurn, Olson, Phillips1, Foot, GK, Moulder, Ryu, Zoran}. These experiments 
have focused primarily on the phenomenon of the metastability of the current-carrying
states. 

Recently, the phenomenon of hysteresis has also been investigated in an annular potential
\cite{hysteresis}. In this experiment a Bose-Einstein condensate of sodium atoms that was 
initially at rest was stirred, and as the rotational frequency of the stirring potential 
increased, the cloud was observed to make a transition to a state with one unit of 
circulation at a critical frequency, $\Omega_1$. On the other hand, in the reverse process 
(i.e., starting with the gas having one unit of circulation and decreasing the frequency 
of the stirrer) the system was observed to return to the state with zero circulation at a 
different critical frequency, $\Omega_2$, which is a clear indication of hysteresis. 

Motivated by the above experiments, we consider here the phenomenon of hysteresis in a 
Bose-Einstein condensed gas of atoms confined in a ring potential \cite{FB, EM, TS, U1, 
U2, BB} as well as the stability of persistent currents \cite{Leg, Yak}.  One of 
the main results of our study is the effect of the finiteness of the atom number on the 
phenomenon of hysteresis and on the stability of the persistent currents. To attack this 
problem we use the method of diagonalization of the many-body Hamiltonian. Contrary to 
the mean-field approximation --- which makes the implicit assumption of a large particle 
number --- the diagonalization approach includes corrections due to a finite number 
of atoms.  In addition, it avoids the assumption of a simple product state for the many-body 
wavefunction that is central to the mean-field approach. As a result, this approach captures 
correlations that are built when the atom number is very low or when the diluteness 
condition is violated. 

We stress that in various recent experiments it has become possible to trap and detect 
very small numbers of atoms, which can even be of order unity see, e.g., Ref.\,\cite{SJ}. 
Indeed, there appears to be a more general tendency in the field of cold atoms to move 
towards the study of small systems.  Interestingly, the vast majority of the theoretical 
studies which have been performed on the superfluid properties of cold atomic gases and 
on the phenomenon of hysteresis assume the opposite limit of large particle numbers, 
since they are based on the mean-field Gross-Pitaevskii approximation. As a result, very 
little is known about the effect of the finiteness of systems with a small number of 
atoms.

In the following we first present our model in Sec.\,II and comment on the phenomena 
of hysteresis and of metastability. Then, we evaluate in Sec.\,III the critical frequencies 
associated with the phenomenon of hysteresis within the mean-field approximation. In 
Sec.\,IV we go beyond the mean-field approximation to consider corrections of order $1/N$ 
(and lower) due to the finiteness of the atom number $N$. In Sec.\,V we investigate the 
same question regarding the critical coupling for metastability and the matrix element 
for the decay rate of persistent currents in a small system. In Sec.\,VI we make contact 
with recent experiments on the phenomenon of hysteresis and of metastability, and finally 
in Sec. VII we present our conclusions.

\section{Model and general considerations}

In the present study we assume one-dimensional motion of bosonic atoms under periodic boundary 
conditions, as in a ring potential. This model is expected to be valid in an annular/toroidal 
trap as long as the interaction energy is much smaller than the energy of the trapping potential 
in the transverse direction. 

If $c_m$ and $c_m^{\dagger}$ are annihilation and creation operators of an atom with angular 
momentum $m \hbar$, the Hamiltonian has the form
\begin{eqnarray}
 {\hat H} = \frac {\hbar^2} {2 M R^2} \sum_m m^2 c_m^{\dagger} c_m 
 + \frac U 2 \sum_{m,n,l,k} c_m^{\dagger} c_n^{\dagger} c_k c_l \, \delta_{m+n, k+l}.
\nonumber \\
\end{eqnarray}
Here $M$ is the atom mass, $R$ is the mean radius of the torus/annulus, $S$ is its cross 
section (in the transverse direction), with $R \gg \sqrt{S}$, and $U = 2 \hbar^2 a/(M R S)$ is 
the matrix element for elastic s-wave atom-atom collisions, with a scattering length $a$.

In analysing the phenomenon of hysteresis and of the metastability of superflow \cite{Leg, Yak}, 
the main feature to be considered is the dispersion relation \cite{FB, EM, TS, U1, U2, BB}, i.e., 
the energy of the system as a function of the angular momentum. Let $E(\ell)$ denote the total 
energy where $\ell \hbar \equiv L \hbar/N$ is the angular momentum per atom and $L \hbar$ 
is the total angular momentum.  According to Bloch's theorem \cite{FB} $E(\ell)$ consists of a 
periodic part plus a quadratic part which comes from the motion of the center of mass.  Thus, one 
needs consider only $0 \le L \le N$ ($0 \le \ell \le 1$); the remainder of the spectrum follows 
trivially as a consequence of Bloch's theorem.

In the absence of interactions $E(\ell)$ consists of straight lines. In the intervals $q \le \ell 
\le q+1$ where $q$ is an integer, $E(\ell)/N = (2 q + 1) |\ell| \hbar^2/(2 M R^2)$ (in what follows
below we assume for simplicity that $\ell \ge 0)$. Obviously, at the end points of each interval 
the first derivative of $E(\ell)$ is discontinuous. In the presence of repulsive/attractive 
interactions these discontinuities remain, while the curvature is negative/positive, respectively. 
Figure 1 shows a schematic picture of the dispersion relation $E(\ell)$ for the repulsive interactions 
which we consider here. Such a spectrum will give rise to hysteresis. If one goes to the rotating 
frame and considers $E_{\rm rot}(\ell)/N = E(\ell)/N - \ell \hbar \Omega$, there are competing local 
minima as the rotational frequency of the trap $\Omega$ is varied. These competing minima give 
rise to discontinuous transitions and thus to hysteresis. The two critical frequencies $\Omega_1$ 
and $\Omega_2$ of the hysteresis loop correspond to the value of the slope of the dispersion relation
$E(\ell)$ for $\ell \to 0^+$ and $\ell \to 1^-$, respectively. The effect of hysteresis is thus a 
generic feature of this problem. On the other hand, for an effective attraction between the atoms, 
hysteresis is absent, since the curvature of $E(\ell)$ is positive, and thus there are no 
discontinuous transitions as the rotational frequency of the trap is varied.

It is convenient to write (in the interval $0 \le \ell \le 1$) the total energy per particle 
$E(\ell)/N$ as \cite{remark}
\begin{eqnarray}
  \frac {E(\ell)} N = \frac {\hbar^2} {2 M R^2} \ell + e(\ell).
\end{eqnarray}
In the case of the non-interacting problem the first term on the right gives the kinetic energy,
and $e(\ell)$ vanishes. Due to Bloch's theorem, $e(\ell)$ is symmetric around $\ell = 1/2$
and a periodic function with a period equal to unity. Expanding $e(\ell)$ for $\ell \to 0^+$, 
$e(\ell) = e(0) + \varepsilon \ell + {\cal O}(\ell^2)$.  This implies that the slope of the 
dispersion relation for $\ell \to 0^+$ is $\hbar^2/(2 M R^2) + \varepsilon$. On the other hand, 
for $\ell \to 1^-$, $e(\ell) = e(0) + \varepsilon (1 - \ell) + {\cal O}(1 - \ell)^2$, and thus 
the slope of the dispersion relation for $\ell \to 1^-$ is $\hbar^2/(2 M R^2) - \varepsilon$.  
In the hysteresis loop it is precisely these slopes that determine the two critical frequencies, 
$\hbar \Omega_1 = \hbar^2/(2 M R^2) + \varepsilon$ and $\hbar \Omega_2 
= \hbar^2/(2 M R^2) - \varepsilon$ as seen in the schematic plot of Fig.\,1.

Therefore, it is crucial to determine the value of $\varepsilon$. Interestingly, the difference
$\hbar (\Omega_1 - \Omega_2)$ is equal to $2 \varepsilon$.  Furthermore, the sign of $\Omega_2$ 
determines the stability of persistent currents. Specifically, the condition $\Omega_2 = 0$ 
represents the critical value of the coupling for metastability of the currents, and metastability 
will be present if $\Omega_2 < 0$.

\begin{figure}
\includegraphics[width=8cm,height=6cm,angle=-0]{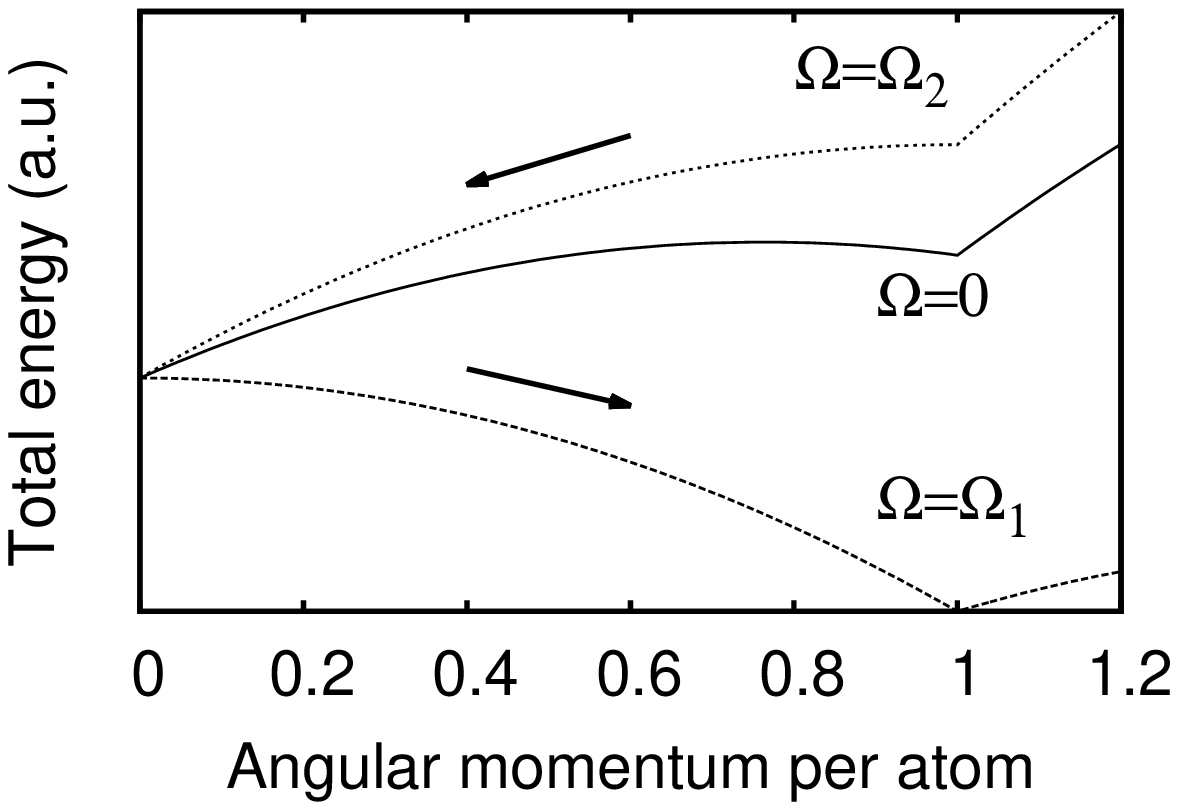}
\includegraphics[width=8cm,height=6cm,angle=-0]{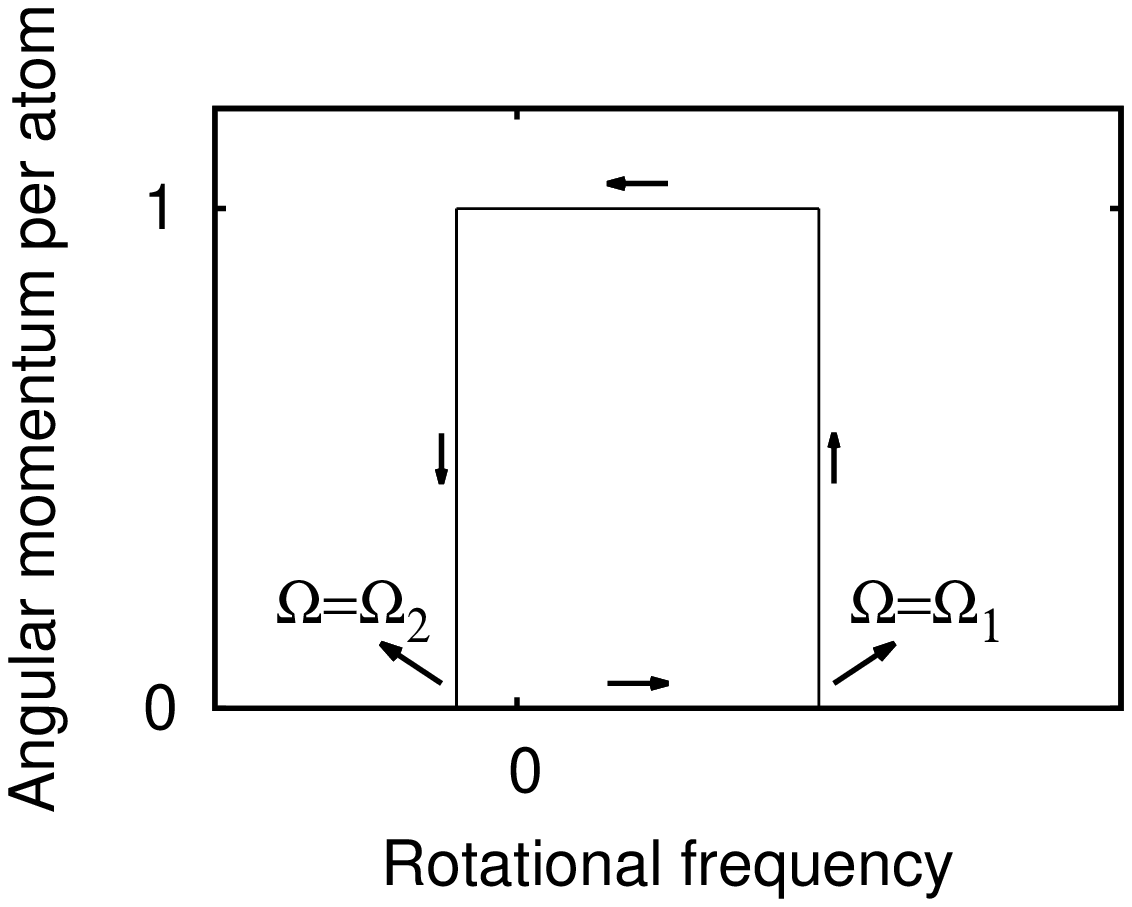}
\caption{Schematic plots showing the energy or, equivalently, the dispersion relation (higher), 
and the corresponding hysteresis loop (lower). In the dispersion relation we show the energy in 
the lab frame, $E(\ell)$, (middle curve) as well as in the rotating frame, $E_{\rm rot}(\ell)/N 
= E(\ell)/N - \ell \hbar \Omega$, for the two critical frequencies $\Omega_1$ and $\Omega_2$ 
for which the slope of $E_{\rm rot}$ vanishes for $\ell \to 0^+$ and $\ell \to 1^-$, respectively. 
The arrows in the upper plot indicate the instability that results from the disappearance of 
the energy barrier; in the lower plot they indicate the hysteresis loop as the rotational
frequency varies.}
\end{figure}

\section{Hysteresis in the mean-field approximation}

We begin with the mean-field approximation and consider the limit $\ell \to 1^-$. One can 
construct a Taylor-series expansion of the energy as a function of the small parameter
$1 - \ell$. Since we are interested in the slope of the dispersion relation, we only need
the linear term in the expansion for the energy. To get that, it suffices to consider only
the dominant state in the order parameter $\Psi$, which is $\phi_1$, with $\phi_m(\theta) 
= e^{i m \theta}/{\sqrt{2 \pi}}$ as well as the neighbouring modes $\phi_0$ and $\phi_2$.
This is due to the fact that there is a cross term in the energy that comes from the 
scattering of two atoms with $m=1$ resulting an atom with $m = 0$ and another on with 
$m = 2$.  This term can be negative and thus lowers the energy \cite{KMP}. Therefore, we 
write the order parameter as
\begin{eqnarray}
  \Psi = c_{0} \phi_{0} + c_1 \phi_1 + c_2 \phi_2,
\label{orpar}
\end{eqnarray}
where the coefficients are real variational parameters and also $|c_1|$ is of order unity, 
while $|c_0|$ and $|c_2|$ are both of order $1 - \ell$. We stress that a completely analogous 
calculation holds for $\ell \to 0$, in which case one should assume that $\Psi = c_{-1} 
\phi_{-1} + c_0 \phi_0 + c_1 \phi_1$. We should also mention that one may work more generally 
with the three states $\phi_{1-\kappa}, \phi_1$ and $\phi_{1+\kappa}$, with $\kappa = 2, 3, 
\dots$, however the fact that the kinetic energy of the states $\phi_m$ scales as $m^2$ 
necessarily implies that $\kappa = 1$.

The coefficients appearing in Eq.\,(\ref{orpar}) must satisfy the normalization condition, 
$c_{0}^2 + c_1^2 + c_2^2 = 1$, and the constraint of fixed angular momentum, $c_1^2 + 2 c_{2}^2 = 
\ell$ or $c_0^2 - c_2^2 = 1 - \ell$. The expectation value of the energy per particle in the above 
state is 
\begin{eqnarray}
 \frac {E} {N \epsilon}  &=&  c_{1}^2 + 4 c_2^2 +  \frac{\gamma}{2} (c_{0}^4 + c_1^4 + c_2^4 
 \nonumber \\ 
&{}& \hspace*{-1em} 
 + 4 c_{0}^2 c_1^2 + 4 c_1^2 c_2^2 + 4 c_{0}^2 c_2^2 - 4 |c_{0}| c_1^2 |c_2|), 
\end{eqnarray}
where $c_0$ and $c_2$ have been assumed to have opposite signs in order to minimize the energy.  
Here, $\gamma/2 = N U/(2 \epsilon) = 2 N a R/S$ is the ratio between the interaction 
energy of the gas with a homogeneous density distribution and the kinetic energy $\epsilon 
\equiv \hbar^2/(2 M R^2)$. After linearisation, the above expression may also be written as 
\begin{eqnarray}
 \frac {E} {N \epsilon} - \frac {\gamma} 2 \approx \ell + 2 c_2^2 + \gamma (|c_0| - |c_2|)^2. 
\end{eqnarray}
Writing $c_0 = \sqrt{1 - \ell} \cosh \theta$, $c_2 = \sqrt{1 - \ell} \sinh 
\theta$, the value of $\theta$ that minimizes the energy is $\theta_0 = (1/4) 
\ln(2 \gamma + 1)$. Therefore, the minimized energy is
\begin{eqnarray}
 \frac {E} {N \epsilon} - \frac {\gamma} 2 \approx \ell + [\sqrt{2 \gamma + 1} - 1] (1 - \ell). 
\end{eqnarray}
The derived value of $\varepsilon$ is thus $\varepsilon/\epsilon = \sqrt{2 \gamma + 1} - 1$ and 
therefore 
\begin{eqnarray}
\Omega_1/\omega = \sqrt{2 \gamma + 1},
\label{om1} 
\end{eqnarray}
while 
\begin{eqnarray}
\Omega_2/\omega = 2 - \sqrt{2 \gamma + 1}, 
\label{omega2}
\end{eqnarray}
where $\omega = \epsilon/\hbar$. We note here that $\Omega_2$ will vanish if $\gamma = 
3/2$.  This is the well-known result for the stability of persistent currents in a single-component 
gas, see, e.g., Ref.\,\cite{OK}.

One can generalize the above results (using Bloch's theorem) in the interval $q \le \ell \le 
q+1$, where 
\begin{eqnarray}
\Omega_1/\omega = 2 q + \sqrt{2 \gamma + 1}, 
\end{eqnarray}
and 
\begin{eqnarray}
\Omega_2/\omega = 2 (q+1) - \sqrt{2 \gamma + 1}. 
\label{omega2g}
\end{eqnarray}
From the last equation it follows trivially that the critical value of the coupling for stability 
of persistent currents (for $\ell = q + 1$) is $\gamma_{\rm} = (2 q + 1) (2 q + 3)/2$, as Bloch's
theorem implies.

\section{Hysteresis beyond the mean-field approximation}

We now examine the same problem beyond the mean-field approximation. To do this, we use
the method of diagonalization of the many-body Hamiltonian. To get some insight, we start
with the truncated space containing the single-particle states $\phi_0$, $\phi_1$, and $\phi_2$ 
[i.e., the states used in Eq.\,(\ref{orpar})]. The eigenstates may be written in the form 
\begin{eqnarray}
  |\Psi_n \rangle = \sum_p c_p^n |p \rangle,
  \label{fock}
\end{eqnarray}
where $n = 0, 1, 2, \dots$ denotes the excited state with index $n$.  Here the states 
$|p \rangle$ are defined as $|0^p, 1^{N - 2 p}, 2^p \rangle$, where the notation 
$|0^{N_0}, 1^{N_1}, 2^{N_2} \rangle$ indicates that $N_0$ atoms occupy the state $\phi_0$, 
etc. Clearly, the states $|p \rangle$ are eigenstates of the number operator and of the 
angular momentum for a system of $N$ atoms with angular momentum $L = N$.  Again, one can 
work more generally with the three states $\phi_{1-\kappa}, \phi_1$ and $\phi_{1+\kappa}$, 
with $\kappa = 2, 3, \dots$, however the corresponding problem becomes block diagonal, with 
the triplet of the states with $\kappa = 1$ giving the slope we are looking for \cite{OK}. 

One can diagonalize the Hamiltonian in this truncated space using the Bogoliubov 
transformation to obtain the eigenvalues ${\cal E}_n(L)$, which are
\begin{eqnarray}
  {\cal E}_n(L=N)/\epsilon &-& \gamma (N-1)/2 = 
  \nonumber \\
  N &-& (\gamma + 1) + \sqrt{2 \gamma + 1} (1 + 2 n).  
\end{eqnarray}
Considering the states $|p' \rangle = |0^{p+1}, 1^{p - 2 m}, 2^{p-1} \rangle$ with $N$ atoms 
and $L = N-2$ units of angular momentum, one can follow the same procedure as before to find 
that
\begin{eqnarray}
{\cal E}_n(L=N-2)/\epsilon  &-& \gamma (N-1)/2 = 
  \nonumber \\
   N - 4 - (\gamma + 1) &+& \sqrt{2 \gamma + 1} (3 + 2 n).
\end{eqnarray}
From the lowest eigenvalues of each of the last two equations it follows that $\Omega_2/\omega 
= 2 - \sqrt{2 \gamma + 1}$, in agreement with the result of the mean-field approximation, 
Eq.\,(\ref{omega2}).

The approach considered above has assumed that $N$ is $\gg 1$, while the expectation value 
of $m$ [in Eq.\,(\ref{fock})] is of order unity. To find the finite-$N$ corrections for the 
critical values of $\Omega_1$ and $\Omega_2$, we have diagonalized the many-body Hamiltonian 
numerically without making any approximations beyond the truncation to some set of single-particle 
states $\phi_m$ with $-m_{\rm max} \le m \le m_{\rm max}$.  Figure 2 shows the result of such a 
calculation for $N = 5$ atoms, $0 \le L \le 10$, $\gamma = N U/\epsilon = 0.5$, and $m_{\rm max} 
= 4$, where we plot a few eigenvalues for each value of $L$. The dispersion relation satisfies 
Bloch's theorem. The fact that the form of this figure is the same as that of the schematic plot 
of Fig.\,1 indicates the presence of hysteresis. We stress that for the small values of $N$ that 
we consider here one can easily reach the Tonks-Girardeau limit. In this limit $\gamma$ is at least 
of order $N^2$. Thus, in order for the mean-field approximation to be valid, $\gamma$ has to be much 
less than $N^2$. 

\begin{figure}
\includegraphics[width=8cm,height=6cm,angle=-0]{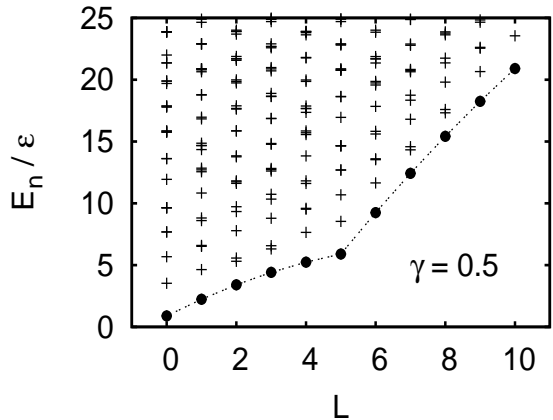}
\caption{The lowest eigenvalues ${\cal E}_n$ of the Hamiltonian for $N = 5$ atoms as a function 
of $L$.  Th calculation was performed for $\gamma = N U/\epsilon = 0.5$, and $m_{\rm max} = 4$.}
\end{figure} 

Having diagonalized the Hamiltonian, we extract the slope of the dispersion relation from 
the difference ${\cal E}_0(L=1) - {\cal E}_0(L=0)$ to determine $\Omega_1$. Finally, by 
varying the atom number, $2 \le N \le 5$ we find that $\Omega_1$ can be approximated as
\begin{eqnarray}
\Omega_1/\omega \approx 1.0953 - 0.8782/N - 0.7513/N^2,
\label{fn} 
\end{eqnarray}
for $\gamma = 0.1$. A subtle point in this calculation is the fact that the interactions 
strength increases with increasing $N$.  This results in a greater depletion of the condensate.  
Thus, in order to extract the critical frequencies associated with the hysteresis, we keep 
$\gamma$ fixed or equivalently allow $U$ to scale like $1/N$.  

In obtaining Eq.\,(\ref{fn}) $m_{\rm max}$ was set equal to $5$. Clearly, $m_{\rm max}$ must 
be sufficiently large so that the fitting parameters have saturated. The differences in these 
parameters due to changing $m_{\rm max} = 4$ to $m_{\rm max} = 5$ are in the seventh, third, 
and second significant figures respectively.  The value of the leading term is remarkably close 
to the value of $\sqrt{1 + 2 \gamma} \approx 1.09544$ found in Eq.\,(\ref{om1}), which is the 
asymptotic value of $\Omega_1$ for $N \to \infty$. Similar calculations for $\gamma = 1$ yield 
\begin{eqnarray}
\Omega_1/\omega \approx 1.7453 - 0.6101/N - 0.1353/N^2.
\label{fnn} 
\end{eqnarray} 
Although the leading term is still reasonably close to $\sqrt{1 + 2 \gamma} \approx 1.73205$,
the agreement is materially worse. This is presumably because of the larger depletion of the
condensate due to the stronger interaction. 

One general observation that emerges from the above analysis is that the effect of the 
finiteness of the system and of the correlations, captured by the method of diagonalization, 
is to decrease the value of $\Omega_1$ from its asymptotic value (and thus to increase the value 
of $\Omega_2$). We comment on this observation in the following section. 

Last but not least, we mention that the value of the angular momentum for which the winding 
number of the order parameter changes is exactly $\ell = 1/2$. In the equivalent language of 
solitary waves \cite{jsmk} the lowest-energy state with this value of the angular momentum  
corresponds to a ``dark" solitary wave (i.e., a solitary wave with a node) which, although dark, 
still has a finite propagation velocity due to the finiteness of the ring \cite{solit, jsmk}.  
Assuming without loss of generality that the center of the solitary wave is located at 
$\theta = \pi$, the real part of the order parameter has a fixed sign. Its minimum value 
(at $\theta = \pi$) vanishes as $\ell \to (1/2)^-$. The imaginary part of the order parameter 
has sinusoidal behaviour and vanishes at $\theta = 0, \pi$, and $2 \pi$. This necessarily 
implies that the net phase change is zero. On the other hand, for $\ell \to (1/2)^+$, the minimum 
value of the real part of the order parameter, which remains $\theta = \pi$, is negative and approaches 
zero from below.  This tiny change in the minimum value of the real part of the order parameter 
from slightly positive to slightly negative is sufficient to change the winding number of the phase. 
We stress that this tiny change can be described perturbatively and, although there is a violent 
rearrangement of the phase of the order parameter, this rearrangement can in no way prevent 
hysteresis.

\section{Metastability of persistent currents in a small system}

The dispersion relation can develop an energy barrier for sufficiently strong and repulsive 
interatomic interactions which separates the state with $L = N$ from the state with $L = 0$
\cite{Leg}. While ${\cal E}_0(L = N)$ will always have a higher energy than ${\cal E}_0(L = 0)$ 
[in fact, ${\cal E}_0(L = N) - {\cal E}_0(L = 0) = N \epsilon$], the state with $L = N$ is then 
metastable. As a result, if the system is prepared in the state $L = N$, it will require an 
exponentially long time for the system to decay since this process must occur via quantum 
tunnelling. Furthermore, the energy and the angular momentum of the gas must be dissipated by 
small non-uniformities in the trapping potential.

In this section we investigate two different questions. The first is the critical value of the 
coupling required for the system to develop an energy barrier with particular concern for 
finite-$N$ effects. The second question is how the matrix element of a symmetry-breaking 
single-particle operator $\Delta V$, that can connect the two eigenstates of lowest energy, 
$|L = N \rangle$ and $|L = 0 \rangle$, depends on the atom number $N$ (for reasons that we 
explain below).

Starting with the first question, according to Eq.\,(\ref{omega2}) the critical value of
$\gamma$ for the existence of a local minimum for $\ell \to 1^-$ is $\gamma_{\rm cr} = 3/2$.
This is an asymptotic result, which does not include finite-$N$ corrections. To find these 
corrections, we choose a fixed value of the atom number $N$ and identify the critical value of 
$U$, $U_{\rm cr}$, which gives a zero slope in the dispersion relation for $\ell \to 1^-$, 
i.e., ${\cal E}_0(L=N) = {\cal E}_0(L=N-1)$. The result of this calculation is given in 
Fig.\,3 where we plot the number of atoms on the $x$ axis and the product $N U_{\rm cr} 
\equiv \gamma_{\rm cr}$ on the $y$ axis, for $m_{\rm max} = 4$.  These results can be 
fit as 
\begin{eqnarray}
  \gamma_{\rm cr} \approx 1.5106 + 0.6020/N + 8.2820/N^2 - 34.8262/N^3 
  \nonumber \\ + 73.3879/N^4.
\end{eqnarray} 
The small deviation of the asymptotic value of $\gamma_{\rm cr}$ in the above expression 
from the expected value of 3/2 is presumably due to the truncation, $m_{\rm max} = 4$, the 
limited number of atoms we have considered, $N \le 10$, and correlations which are absent 
in the calculation within the mean-field approximation. Interestingly, as seen from Fig.\,3,
the value of $\gamma_{\rm cr}$ for a finite number of atoms is higher than 3/2. Since this
is determined by the slope ${\cal E}_0(L=N) - {\cal E}_0(L=N-1)$, we conclude that the
correlations which are captured within the present approach (but are absent within the
mean-field approximation) lower the energy of the state with $L=N-1$ more than that of the 
state with $L=N$.  Thus, a higher value of $\gamma$ is necessary to stabilize the currents
in the state with $L = N$. The same mechanism which increases $\gamma_{\rm cr}$ is also
responsible for the decrease (increase) of $\Omega_1 (\Omega_2)$ found in the previous 
section.

We turn now to the second question regarding the decay rate of the persistent current. 
In order for the energy barrier (which develops for sufficiently strong interatomic interactions)
to prevent the decay of the currents and render them metastable with an exponentially long decay 
time, the matrix element of any symmetry-breaking single-particle operator $\Delta V$ connecting 
the states $|L = N \rangle$ and $|L = 0 \rangle$ must be vanishingly small \cite{note}. Otherwise 
the presence of the energy barrier becomes irrelevant and the currents will decay. 

To investigate this problem, we consider a single-particle operator $\Delta V = V_0 \sum_{i=1}^N 
\delta(\theta_i)$, which is a sum of delta function potentials intended to mimic irregularities in 
the trap \cite{TS}. This potential breaks the axial symmetry of the Hamiltonian and induces 
transitions between the two states $|L = N \rangle$ and $|L = 0 \rangle$. We thus evaluate the 
matrix element $\langle L = N |\Delta V |L = 0 \rangle$, making use of the lowest-energy states 
$|L = 0 \rangle$ and $|L = N \rangle$ that we get from the diagonalization of the axially-symmetric 
Hamiltonian. Clearly the only terms which give a nonzero contribution to this matrix element are 
those that raise the angular momentum by $L = N$ units when acting on $| L = 0 \rangle$, 
\begin{eqnarray}
  \langle L = N |\Delta V |L = 0 \rangle = V_0 \sum_{n}
  \langle L = N | a_n a_{n+N}^{\dagger} |L = 0 \rangle.
  \nonumber \\
\end{eqnarray}
In the absence of interactions, when all the atoms are in the single-particle state $\phi_1$ and $
\phi_0$, respectively, this matrix element vanishes for all $N > 1$.  This is also the case in the 
mean-field approximation. To get a non-vanishing matrix element it is necessary to consider non-zero 
interactions that deplete the condensate and a finite number of atoms. 

\begin{figure}
\includegraphics[width=8cm,height=6cm,angle=-0]{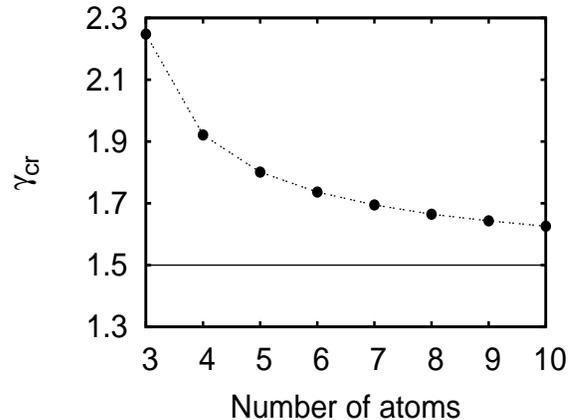}
\caption{The value of $\gamma_{\rm cr}$ obtained  with the method of diagonalization as a function 
of $N$ for $m_{\rm max} = 4$. The horizontal line shows the asymptotic value of 
$\gamma_{\rm cr} = 3/2$.}
\end{figure}

\begin{figure}
\includegraphics[width=8cm,height=6cm,angle=-0]{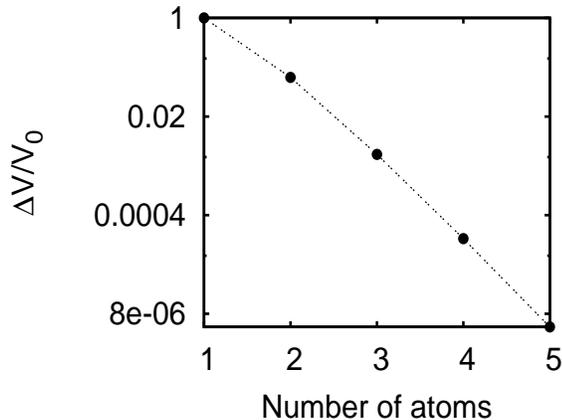}
\caption{The matrix element of the operator $\Delta V$ between the states with $|L = N \rangle$
and $|L = 0 \rangle$, $|\langle L = N |\Delta V |L = 0 \rangle/V_0|$, as function of the atom
number $N$, for a fixed value of $\gamma = g N = 0.1$. Here the states $|L = 0 \rangle$ and 
$|L = N \rangle$ have been evaluated for $m_{\rm max} = 5$.}
\end{figure}

\begin{figure}
\includegraphics[width=8cm,height=6cm,angle=-0]{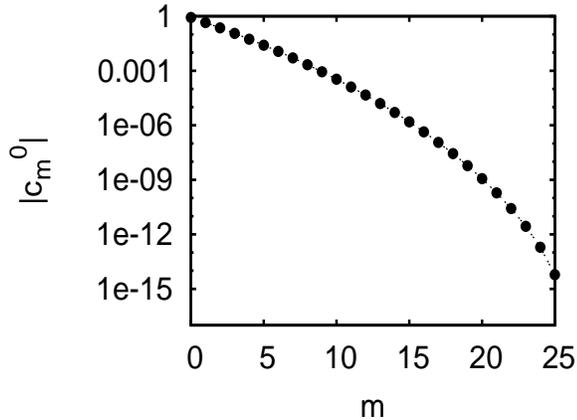}
\caption{The amplitudes $|c_m^0|$ which appear in Eq.\,(\ref{fock}) as function of the index 
$m$, for $N = 50$ atoms, $L = 0$, $\gamma = 5$, and truncation within the single-particle states 
$\phi_{-1}$, $\phi_0$, and $\phi_1$.} 
\end{figure}

Figure 4 shows the value of $|\langle L = N |\Delta V |L = 0 \rangle/V_0|$ as function of
$N$.  Again, we keep $\gamma = g N$ fixed for the reasons stated above.  Here, we have 
chosen $\gamma = g N = 0.1$, while the states $|L = 0 \rangle$ and $|L = N \rangle$ have been 
evaluated for $m_{\rm max} = 5$. As seen from this plot, this matrix element shows an
exponential decay as function of $N$. 

To get an understanding of this decay we recall that the operator $\Delta V$ excites atoms,
increasing their angular momentum by $N$ units. Furthermore, the amplitudes $c_m$ in the expression 
of Eq.\,(\ref{fock}) decay very rapidly with $m$, as seen in Fig.\,5 for $N = 50$ atoms with a rate 
that does not depend on $N$. This is a more general result that also holds in more extended spaces.  
The fact that the amplitudes of the states contributing to $|L = 0 \rangle$ and $|L = N \rangle$ 
decrease rapidly as one moves away from $|0^N \rangle$ and $|1^N \rangle$ along with the nature of 
$\Delta V$, which induces single-particle excitations by $N$ units of angular momentum, combine 
to make this decay matrix element extremely sensitive to $N$.

Thus, the main result of this section is, quite generally, that a combination of sufficiently 
strong interatomic interactions and a finite number of atoms enhances the size of the matrix 
element and thus reduces the timescale that is associated with the decay rate of the persistent 
currents. This result may be interesting to explore experimentally in small systems with an 
interaction whose strength can be tuned.

\section{Connection with the experiments on hysteresis and metastability}

In order for our assumption of one-dimensional motion to be valid, the interaction energy
must be much smaller than the quantum of energy associated with the motion of the atoms
in the transverse direction (or, equivalently, the coherence length must be much larger
than the transverse dimensions of the annulus/torus). However, this assumption is violated 
under current typical conditions, and thus the motion is not quasi-one-dimensional. 

For example, in the experiment of Ref.\,\cite{hysteresis}, where $^{23}$Na atoms were used, 
the chemical potential is $\mu/\hbar \approx 2 \pi \times 1.7$ kHz, while the frequencies of 
the annular-like trapping potential (in the transverse direction) are $\omega_1 \approx 472$ 
Hz and $\omega_2 \approx 188$ Hz. (As a result, it has been argued that vortex-antivortex pairs 
form in this experiment.) Thus, it is not possible to make neither a quantitative nor a
qualitative comparison of the present theory and the experiment of Ref.\,\cite{hysteresis}. 
An investigation of this problem using a more realistic model is underway and will be described 
in a future publication.

If one wants nonetheless to get an estimate for the critical frequencies of hysteresis for 
the parameters of Ref.\,\cite{hysteresis} using the present theory, it follows for a radius 
of $R \approx 19.5$ $\mu$m, that $\omega = \hbar/(2 M R^2) \approx 3.6$ Hz. Given that $a 
\approx 28$ \AA, $N \approx 4 \times 10^5$, and $S = \pi a_1 a_2$ with $a_i = \sqrt{\hbar/
(M \omega_i)}$, i.e., $a_1 \approx 2.42$ $\mu$m and $a_2 \approx 3.83$ $\mu$m, the 
dimensionless parameter $\gamma = 2 N a R/S$ has the value $\gamma \approx 1500.0$. It then
follows from Eqs.\,(\ref{om1}) and (\ref{omega2}) that $\Omega_1 \approx 197.2$ Hz and 
$\Omega_2 \approx - 190.0$ Hz. Clearly, these large frequencies (as compared to the observed 
frequencies, which are on the order of 10 Hz) are due to the very large value of $\gamma$, 
which is the ratio between the interaction energy of a homogeneous cloud with a density $n_0 
= N/(2 \pi R S)$ and the kinetic energy associated with the motion in the ring, $\hbar^2/
(2 M R^2)$.

It is also interesting to make estimates for the case where the motion is quasi-one-dimensional. 
Consider, for example, the case where the experimental conditions are identical to those of 
Ref.\,\cite{hysteresis} but where the number of atoms is reduced by, e.g., a factor of $4 
\times 10^4$ to the value $N = 10$. This would reduce the interaction energy to the extent 
that the conditions for one-dimensional motion would be fulfilled. This reduction in $N$ 
would also reduce the value of $\gamma$ to $\approx 0.04$. The corresponding critical 
frequencies would become $\Omega_1 \approx 3.7$ Hz and $\Omega_2 \approx 3.5$ Hz. While the 
difference between $\Omega_1$ and $\Omega_2$ is small, $\approx 2 \gamma \omega$, it would 
still be of interest to investigate their dependence on $N$, which, according to the results 
of Sec.\,IV, is $1/N$ to leading order. 

It would also be interesting to investigate the effect of finite system size on the critical 
value for stability of the persistent currents in such small systems. According to the results 
of Sec.\,V, the value of $\gamma_{\rm cr}$ also scales as $1/N$ to leading order. Last but not 
least, the decay time of the currents would show a much more rapid -- and thus more pronounced 
-- decrease as $N$ decreases. 

\section{Conclusions}

In the present study we have investigated the phenomenon of hysteresis and of metastability in 
a Bose-Einstein condensed cloud of atoms which are confined in a ring potential. Interestingly, 
this problem has recently been examined experimentally \cite{hysteresis}, while many other 
experiments have focused on the question of persistent currents in such topologically 
nontrivial potentials \cite{Kurn, Olson, Phillips1, Foot, GK, Moulder, Ryu, Zoran}.

In the phenomenon of hysteresis the main question is the evaluation of the critical frequencies.
As we have shown, in a purely one-dimensional system these two frequencies are related as a 
consequence of Bloch's theorem. Further, we have evaluated those both within the mean-field 
approximation and beyond mean field (i.e., by numerical diagonalization of the many-body Hamiltonian)
in order to determine finite-$N$ corrections. 

We have also performed calculations of the critical coupling for the metastabiliity of superflow and 
of the matrix element associated with the decay rate in a finite system of atoms. As we have argued, 
the depletion of the condensate due to the interaction combined with the finiteness of the atom 
number can cause the decay rate to increase exponentially with decreasing $N$. Thus, the general 
tendency is that the finiteness of a system makes the supercurrents more fragile, in the sense that 
it increases the decay rate of the currents, and it also increases the critical coupling for 
metastability. 

Given the recent experimental activities on the problems of hysteresis and of metastability, and
also given the more general tendency in the community of cold atoms to move to small systems (i.e., 
systems with a small atom number $N$) the present results, which we believe are of theoretical 
interest, may will become experimentally relevant in the near future.

\acknowledgements

This project is implemented through the Operational Program ``Education and Lifelong Learning", 
Action Archimedes III and is co-financed by the European Union (European Social Fund) and Greek 
national funds (National Strategic Reference Framework 2007 - 2013).


\begin{thebibliography}{99}
\bibitem{remark} We stress that this expression is consistent with Bloch's theorem, since
the energy may also be written as $E(\ell)/N = [\hbar^2/(2 M R^2)] [\ell^2 + (\ell^2 
- \ell)] + e(\ell)$, where the last three terms are periodic.
\bibitem{note} There is another way for the currents to decay, via high-order perturbation 
theory, however this process is also highly suppressed for weak irregularities.
\end{thebibliography}
\end{document}